# The evolution of galaxies from primeval irregulars to present-day ellipticals

Masao Mori[1,2] & Masayuki Umemura[3]

[1]*Department of Physics and Astronomy, University of California, Los Angeles, California 90095-1547, USA*

[2]*Institute of Natural Sciences, Senshu University, Kawasaki, Kanagawa 214-8580, Japan*

[3]*Center for Computational Sciences, University of Tsukuba, Tsukuba, Ibaraki 305-8577, Japan*

**The current understanding of galaxy formation is that it proceeds in a 'bottom up' way, with the formation of small clumps of gas and stars that merge hierarchically until giant galaxies are built up[1,2]. The baryonic gas loses the thermal energy by radiative cooling and falls towards the centres of the new galaxies, while supernovae (SNe) blow gas out[3,4]. Any realistic model therefore requires a proper treatment of these processes, but hitherto this has been far from satisfactory[5]. Here we report an ultra-high-resolution simulation that follows evolution from the earliest stages of galaxy formation through the period of dynamical relaxation. The bubble structures of gas revealed in our simulation ($< 3 \times 10^8$ years) resemble closely the high-redshift Lyman α emitters (LAEs)[6,7]. After $10^9$ years these bodies are dominated by stellar continuum radiation and look like the Lyman break galaxies (LBGs)[8,9] known as the high-redshift star-forming galaxies at which point the abundance of elements heavier than helium ("metallicity") appears to be solar. After $1.3 \times 10^{10}$ years, these galaxies resemble present-day ellipticals.**

To explore the early evolution of galaxies, the coupling of the dynamics and the chemical evolution through star formation and SN feedback should be treated properly[10,11]. Especially, it is crucial to resolve accurately the thermalization of the kinetic energy released by multiple SNe. We present an ultra-high-resolution ($1024^3$ fixed Cartesian grids) hydrodynamic simulation coupled with the collisionless dynamics for dark matter particles and star particles, which is solved by an *N*-body method. The simulation pursues the early evolution ($< 2\times10^9$ years) of a proto-galaxy as an assemblage of sub-galactic condensations with a mass of $5.0\times10^9$ $M_\odot$ building up the total mass of $10^{11}$ $M_\odot$. The details of the numerical procedures are described in the Supplementary Information.

Figure 1 shows the result for the time sequence of the star formation, gas dynamics, and the chemical enrichment. In the first $10^8$ years, stars form in high-density peaks within sub-galactic condensations and the burst of star formation starts. Then, massive stars in the star forming regions explode as SNe one after another. The gas in the vicinity of SNe is quickly enriched with ejected metals, but a large amount of gas still retains low metal abundance. Consequently, the metallicity distribution becomes highly inhomogeneous on kpc scales, where gas enriched as $-5 \leq [O/H] \leq -1$ coexists with virtually primordial gas. Since the density of the interstellar medium is lower in the outer regions of sub-galactic condensations, the expansion of hot bubbles is accelerated there. At $3\times10^8$ years, SN-driven shocks collide with each other to generate super-bubbles of ~50 kpc, and the surrounding high-density, cooled ($10^4$K) shells. The dense shells undergo hydrodynamic instabilities induced by radiative cooling, eventually fragmenting into cold filaments and blobs. New stars are born in the enriched gas and subsequent SNe again eject heavy elements. The hot bubbles expand further by continual SNe, and the shells sweep up the partially enriched ambient gas. The gas density in dense shells increases owing to the efficient radiative cooling mainly through collisional excitation of neutral hydrogen. After $5\times10^8$ years, the hot bubbles blow out into the intergalactic space. The rightmost panels show the structure at $10^9$ years. By this stage, the interstellar medium is recycled repeatedly. Eventually, some amounts of cool, dense filaments are left at the centre. But, most of volume is filled with rarefied gas (~$10^{-4}$ cm$^{-3}$) that has intermediate temperature ($10^{4.5} \leq T\,(\mathrm{K}) \leq 10^{6.5}$). At this epoch, the mixing of heavy elements is nearly completed.

Newly born stars trace the mixing history of the heavy elements well, because they inherit the metal abundance of gas. In Figure 2, the star formation epoch is shown as a function of the oxygen abundance of newly formed stars. It is clearly seen that, before $10^8$ years, there is considerable variance in the oxygen abundance ($-5 \leq [\mathrm{O/H}] \leq -1$), reflecting a very inhomogeneous distribution of enriched gas. After $10^8$ years, the merger of sub-galactic condensations promotes the mixing of heavy elements. Finally, the almost complete recycling of interstellar matter erases the inhomogeneities of metal abundance. As a result, the oxygen abundance of stars converges to $-0.3 \leq [\mathrm{O/H}] \leq 0.2$ with small dispersion. It is worth noting that the metal abundance is already at the level of solar abundance at $10^9$ years.

In Figure 3, the spectral energy distribution (SED), the surface brightness distributions, and the star formation history are shown. The star formation rate increases in $5\times10^7$ years, and reaches a peak of about 40 $M_\odot$ yr$^{-1}$ around $1.5\times10^8$ years. The burst of star formation continues until $3\times10^8$ years. Then, the star formation activity gradually diminishes down to a few $M_\odot$ yr$^{-1}$ after $10^9$ years because SN-driven winds have removed any remaining cold gas from the sub-galactic fragments. As seen in the SED, at the earliest stages of less than $3\times10^8$ years, the Lyman α (Lyα) emission is conspicuous, which comes from high-density cooling shells and its luminosity is more than $10^{43}$ erg s$^{-1}$. The Lyα luminosity perfectly matches that observed in LAEs[12,13]. This result suggests that LAEs could correspond to an early SN-dominated phase before $3\times10^8$ years. Among theoretical models for LAEs[7,14,15], the present multiple SN model is distinctive in the bubbly structure. In Figure 4, the narrow-band image of extended LAE observed by Matsuda *et al.*[12] is compared to the distribution of the Lyα emission of the simulated galaxy at $2\times10^8$ years. We find that the physical extent of ~100 kpc and the bubbly structure produced by multiple SNe are quite similar to the observed features in Lyα surface brightness distribution of this LAE.

After $3\times10^8$ years, the Lyα luminosity quickly declines to several $10^{41}$ erg s$^{-1}$, since the emission from cooling gas decreases immediately owing to the leak of explosion energy through the blowouts of super-bubbles. Then, the SED becomes dominated by stellar continuum emission. The galaxy in this phase is featured with diffuse, asymmetric structures, and outflows of 100~500 km s$^{-1}$. The total mass of long-lived stars is $9.3\times10^9\,M_\odot$, and the mass of $1.5\times10^9\,M_\odot$ is involved in the outflows at $z$=3. These features look quite similar to those observed for LBGs[16,17]. The low-ionization interstellar absorption lines observed in LBGs are blueshifted by hundreds km s$^{-1}$ relative to systemic velocities and Lyα lines are redshifted to the same degree. Furthermore, the strong metal absorption lines observed in the spectra of LBGs indicate that their star formation events must have been preceded by an earlier starburst. The excess of absorption-line system with large CIV column density in spectra of background quasars near LBGs is interpreted as further evidence for chemical enrichment of the intergalactic medium due to the SN-driven outflows. Recently, the X-ray luminosity[18,19] at 2.0-8.0 keV for LBGs has been found to be ~$10^{41}$ erg s$^{-1}$. In the present simulation, the X-ray luminosity at the same energy range changes from $10^{42}$ erg s$^{-1}$ at $3\times10^8$ years to ~$10^{41}$ erg s$^{-1}$ around $10^9$ years. The LBG metallicity appears to be the solar abundance for massive systems[20]. In the light of such properties, the simulated post-starburst galaxy with the age of $10^9$ years can correspond to LBGs. Thus, it is implied that LBGs are the subsequent phase of LAEs.

The long-term dynamical evolution of the model galaxy was studied with an *N*-body simulation containing one million particles. As a result, it is found that the assembly of sub-condensations and the virialization of the total system are almost completed in $3\times10^9$ years, so that the system becomes in a quasi-equilibrium state. The resultant stellar system forms a virialized, spheroidal system. The right upper panel in Fig. 3 shows the projected surface brightness distributions in the *U, B, V*, and *K* bands at $1.3 \times 10^{10}$ years (redshift $z=0$) assuming the passive evolution (no further star formation). They have a large central concentration that well accords with de Vaucouleurs' $r^{1/4}$ profile[21], which is commonly found in nearby elliptical galaxies[11,22]. The resultant absolute magnitude in blue band and visual band are $M_B$=-17.2 mag and $M_V$=-18.0 mag, respectively. The colour *U-V*=1.15 and *V-K*=2.85 are consistent with the colour-magnitude relation of elliptical galaxies in Coma cluster of galaxies[22]. Furthermore, the combination of the surface brightness, the effective radius $r_e$=3.97 kpc, and the central velocity dispersion $\sigma_0$=133 km s$^{-1}$ is on the fundamental plane of elliptical galaxies within their scatters (the fundamental plane is the relationship among these three parameters derived for nearby elliptical galaxies)[23,24]. Thus, it is suggested that LBGs evolve into elliptical galaxies through purely collisionless dynamical evolution.

**Supplementary Information** is linked to the online version of the paper at **www.nature.com/nature**.

**Acknowledgments** We thank Y. Matsuda and his collaborators for use of the observational data by Subaru Telescope, and are grateful to M. Rich, M. Malkan, I. Saviane, Y. Yoshii, R. Ellis and the anonymous referees for many helpful suggestions. M.M. thanks University of California, Los Angeles for its hospitality. The simulations were performed with the Earth Simulator at the JAMSTEC, the SPACE at Senshu University, and the computational facilities including CP-PACS at CCS in University of Tsukuba. M. M. is grateful to the JSPS and M. U. to the Ministry of Education, Culture, Sports, Science, and Technology.

**Competing interests statement**    The authors declare that they have no competing financial interests.

**Correspondence** and requests for materials should be addressed to M.M. (mmori@isc.senshu-u.ac.jp).

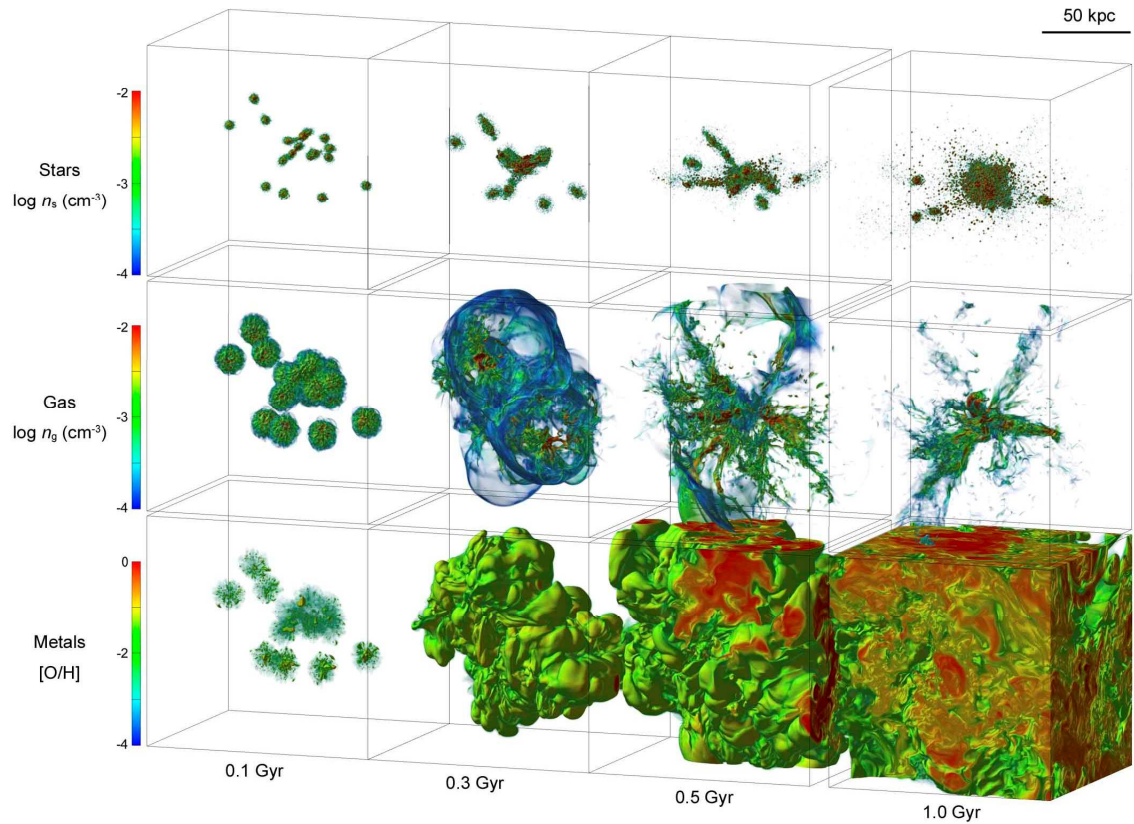

**Figure 1. The first 1Gyr ($10^9$ years) simulation of a proto-galaxy with the total mass of $10^{11}$ $M_\odot$ (solar mass).** The overdensity region of this mass-scale decouples from the cosmic expansion at redshift $z = 7.8$ at a radius of 53.7 kpc (1 kpc=3,260 light-years), where the initial conditions are set up. The mass of gaseous matter is $1.3\times10^{10}$ $M_\odot$ initially. The angular momentum is provided by a spin parameter of $\lambda = 0.05$[25]. Here, we assume the ΛCDM cosmology with $\Omega_M$ = 0.3, $\Omega_b$ = 0.04, and $\Omega_\Lambda$ = 0.7, where $\Omega_M$ is the matter density, $\Omega_b$ baryon density, and $\Omega_\Lambda$ the cosmological constant. The Hubble constant is assumed to be $H_0$ = 70 km s$^{-1}$ Mpc$^{-1}$. The density profiles in sub-galactic dark halos are given by the Navarro-Frenk-White profile[26] and these condensations are distributed randomly within the galaxy-scale overdensity. Radiative cooling for the gaseous component is calculated using the cooling function for an optically-thin,

collisionally-ionized gas[27]. Stars are assumed to form in gravitationally unstable cooled regions with Salpeter's initial mass function[28], at a rate that is inversely proportional to the local free-fall time[10,22]. Stars more massive than 8 $M_\odot$ explode as type II SNe with the explosion energy of $10^{51}$ ergs, and eject synthesized heavy elements. The evolution is shown by the spatial distributions of the stellar density, the gas density, and the oxygen abundance defined by [O/H]=$\log_{10}(N_O/N_H)$-$\log_{10}(N_O/N_H)_\odot$ for gas, where $N_O$ and $N_H$ are the number densities of oxygen and hydrogen, respectively. Each simulation box has a physical size of 134 kpc and the spatial resolution is 0.131 kpc. This is comparable with typical size of super-bubbles observed in the local universe. Both the number density of stellar component and that of gas component range from $10^{-4}$ cm$^{-3}$ to $10^{-2}$ cm$^{-3}$ and the gas metallicity ranges from -4 to 0.

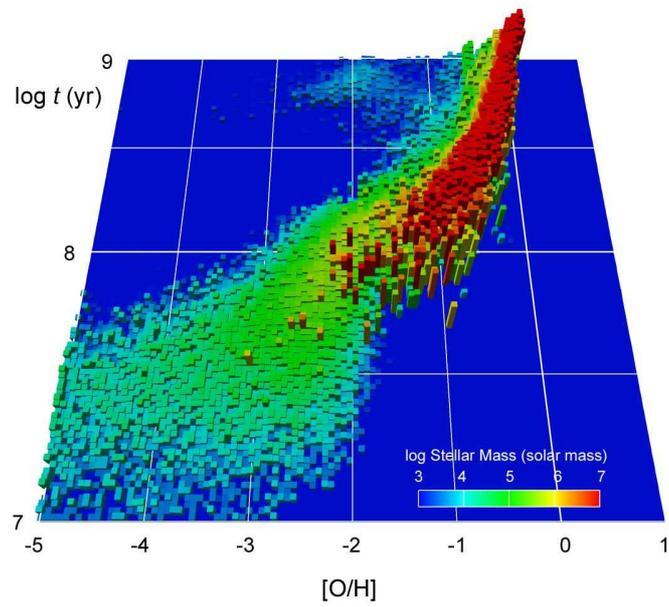

**Figure 2. The formation epochs of stars in the units of year as a function of stellar oxygen abundance [O/H].** The colour-coded histogram denotes the stellar mass in the units of solar mass in logarithmic scales.

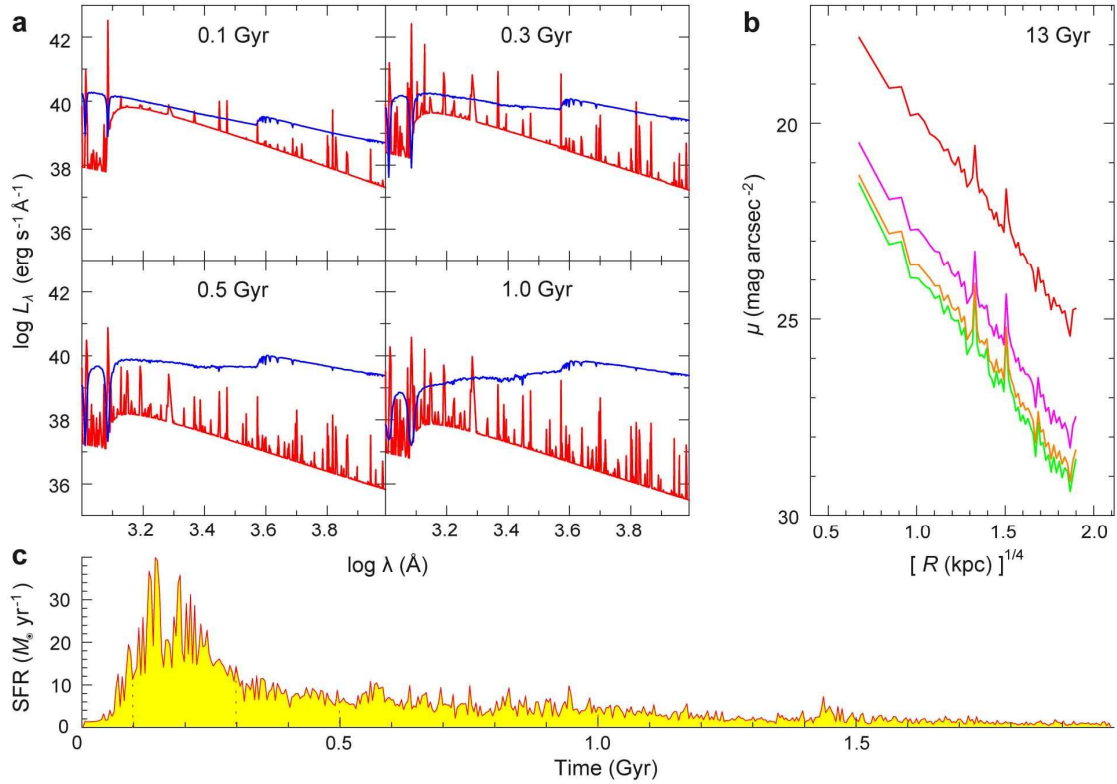

**Figure 3. Emissions and star formation history. a,** predicted spectral energy distribution (SED) of the emission from the simulated galaxy. The emission properties of the gas components are calculated for an optically-thin, collisionally-ionized gas using MAPPINGIII code[27] (red lines), and those of the stellar components are calculated using the evolutionary stellar population synthesis code PÉGASE[30] (blue lines). In practice, to obtain the SED, we sum up the SED of each grid point for the gas components and each star particle for the stellar components. The absolute luminosities of Lyα line emission, where the wavelength is 1216Å, are $2.0\times10^{43}$ erg s$^{-1}$, $1.6\times10^{43}$ erg s$^{-1}$, $4.6\times10^{41}$ erg s$^{-1}$, and $2.3\times10^{41}$ erg s$^{-1}$ at the elapsed time of 0.1 Gyr, 0.3 Gyr, 0.5 Gyr, and 1 Gyr, respectively. **b,** projected distribution of surface brightness at 13 Gyr for our simulation run. Solid lines from bottom to top are the surface brightness in the *U*, *B*, *V*, and *K* bands,

respectively, plotted against a quatic root of the radius (in the units of kpc). **c,** star formation rate (in the units of $M_\odot$ yr$^{-1}$) as a function of time (in the units of Gyr).

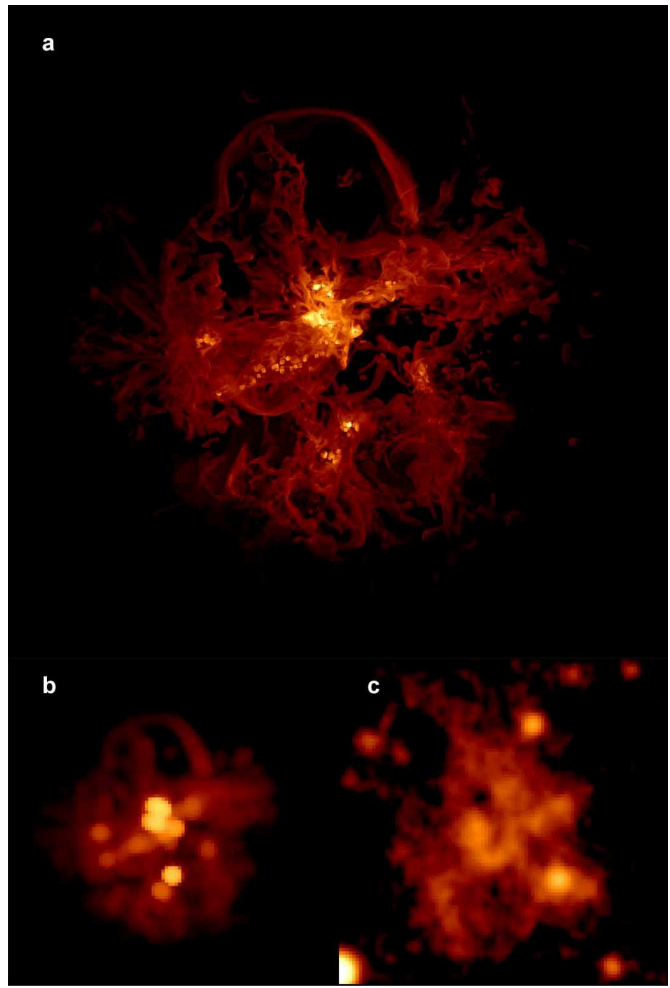

**Figure 4. Comparison of the simulation and observation.** All panels are 154 kpc square. **a,** The projected distribution of the Lyα emission from gas component for the simulated galaxy at the $2.0\times10^8$ years. **b,** The simulation result smoothed with a Gaussian kernel with a FWHM of 7.6 kpc, which corresponds to 1.0" at the redshift $z$=3.09, the same resolution (100 pixels x 100 pixels) as the observation[12]. **c,** Narrow-band image of extended Lyman α emitter ``LAB1'' taken with the Subaru Telescope in the SSA22 field at the redshift $z$=3.09[12].

**Supplementary Methods**

# Galactic evolutionary path from primeval irregulars to present-day ellipticals

Masao Mori[1, 2] and Masayuki Umemura[3]

[1]*Department of Physics and Astronomy, University of California, Los Angeles, California 90095-1547, USA*

[2]*Institute of Natural Sciences, Senshu University, Kawasaki, Kanagawa 214-8580, Japan*

[3]*Center for Computational Sciences, University of Tsukuba, Tsukuba, Ibaraki 305-8577, Japan*

## 1. Introduction

The traditional models of galactic chemical evolution based on a scheme of stellar population synthesis have accounted for the observed colour-magnitude relation of elliptical galaxies[1, 2]. Such model has been grounded on two basic assumptions, that is, the prompt thermalization of the kinetic energy released by supernova (SN) explosions, and the instantaneous, perfect mixing of synthesized heavy elements. However, these assumptions are obviously problematic. The energy input and metal ejection by SN explosions are likely to proceed in an inhomogeneous fashion[3-5]. Thus, simulations that can resolve SN remnants are required to properly model the chemical evolution of galaxies.

On the other hand, the hydrodynamic models of elliptical galaxies are motivated by the observation that elliptical galaxies have very little interstellar medium[6]. The models predict that SN explosions heat the gas in elliptical galaxies and drive galactic winds[7, 8], which blow out of the galaxies. The recent developments in computer technologies and numerical methods have made it possible to simulate the multi-dimensional dynamical or chemodynamical evolution of galaxies including the effect of star formation and SN feedback[9-23]. A dynamical, chemical, and spectrophotometric scheme has become a powerful tool to explore the formation and evolution of galaxies[24-34].

In this Supplementary Information, we describe the initial condition of the present model and our dynamical, chemical, and spectrophotometric scheme to explore the formation and evolution of elliptical galaxies including star formation and SN feedback.

## 2. Initial Condition

In this Letter, by performing an ultra-high-resolution simulation, we attempt to trace the early history of dynamical and chemical evolution of an elliptical galaxy. The hydrodynamic processes are modelled with $1024^3$ fixed Cartesian grids for a proto-galaxy with the total mass of $10^{11}$ $M_\odot$. The total mass of gaseous matter is assumed to be $1.3\times10^{10}$ $M_\odot$ initially.

Following a standard scenario of the ΛCDM universe ($\Omega_M = 0.3$, $\Omega_\Lambda = 0.7$, $\Omega_b = 0.04$, and a Hubble constant of $H_0 = 70$ km s$^{-1}$ Mpc$^{-1}$), we consider a proto-galaxy originating from a CDM density fluctuation with $\delta M/M$ equal to $\nu$ times the rms value $\sigma$. The fluctuation is normalized to unity for a spherical top-hat window of comoving radius 8 Mpc with a bias parameter of $b=1$. Here, we consider the evolution of a relatively isolated $2\sigma$ density fluctuation in the ΛCDM universe with sub-galactic dark halos originating from $2\sigma$ density fluctuations. In this case, sub-galactic dark matter halos with the mass of $\sim 5\times10^9$ $M_\odot$ collapse and virializes at the redshift $z=7.8$. On the other hand, a dark matter halo with the mass of $10^{11}$ $M_\odot$ decouples from the cosmic expansion and begins to collapse at the same redshift. In the light of these points, we simply model a proto-galaxy with the total mass of $10^{11}$ $M_\odot$ as an assemblage of sub-galactic condensations with a mass of $5\times10^9$ $M_\odot$.

The virialization of the CDM halos through the hierarchical clustering has been investigated with $N$-body simulations, and systematic simulations have predicted a 'universal density profile' known as the Navarro-Frenk-White profile[35]. On the other hand, a high-resolution simulation showed that the CDM halos have a steeper central cusp than quoted above[36, 37]. The resulting structure of the CDM halos depends on the number of particles used in the simulation. Here, we assume that sub-galactic condensations have the Navarro-Frenk-White density profile:

$$\rho_{\rm NFW}(r) = \frac{\rho_c \delta_c}{(r/r_s)(1+r/r_s)^2}, \quad (1)$$

with

$$\delta_c = \frac{200}{3} \frac{c^3}{\ln(1+c) - c/(1+c)}, \quad (2)$$

where $\rho_c$ is the critical density, $r_s$ is a scale radius, $\delta_c$ is a characteristic density, and $c$ is a concentration parameter. We set to the concentration parameter to be $c=4$. The gas has the homogeneous distribution within a virial radius of the sub-galactic condensation.

Then the twenty sub-galactic condensations with the mass of $5\times10^9$ $M_\odot$ are distributed randomly within a turn-around radius of 53.7 kpc. The angular momentum is provided by a uniform rotation characterized by a spin parameter of $\lambda = 0.05$[38].

## 3. Numerical Method

Our simulation uses a hybrid *N*-body/hydrodynamic code that is applicable to a complex system consisting of dark matter, stars and gas. The energy equation is solved with radiative cooling processes. The gas is allowed to form stars when cooling instability occurs, and the energy feedback from SNe is included. The chemical enrichment and the photometric evolution of the system can be also simulated by this code.

Mori, Umemura and Ferrara[5] have used an exceedingly high resolution simulation to model a forming galaxy undergoing multiple SN explosions and found that their models could reproduce the typical Lyα luminosities observed in Lyα Blobs[39]. Also, the bubbly features[40] recently discovered in some LABs are quite similar to the structure predicted in their simulation. However, they just considered the first $5\times10^7$ years of the galaxy. Therefore, they assumed the fixed gravitational potential of dark matter halo without the dynamics of dark matter. Thus, how the subsequent evolution proceeds and how it is related to Lyman break galaxies or present-day galaxies are not studied.

In this Letter, we calculate the hydrodynamic evolution over $2\times10^9$ years with the dynamics of the dark matter halo and also pursue the subsequent stellar dynamical evolution over $1.3\times10^{10}$ years. Thus, the present analysis allows us to explore the physical connection among Lyman α emitters[41-46], Lyman break galaxies[47-55], and present-day elliptical galaxies[56-59], by comparing the simulation directly to the observations of those galaxies.

### 3.1 Collisionless dynamics

The equation of motion for collisionless particles (dark matter or stars) is calculated by

$$\frac{d\mathbf{r}_i}{dt} = \mathbf{v}_i, \quad \frac{d\mathbf{v}_i}{dt} = -G\sum_{j\neq i}\frac{m_j}{(|\mathbf{r}_i - \mathbf{r}_j|^2 + a^2)^{3/2}}(\mathbf{r}_i - \mathbf{r}_j), \quad (3)$$

where $m_i$ is a mass, $\mathbf{r}_i$ is a position vector, and $\mathbf{v}_i$ is a velocity vector of a collisionless particle, $G$ is the gravitational constant, and $a$ is a softening parameter. The softening procedure is known as 'Plummer softening' since it replaces the potential of each point with that of Plummer potential[60]. Here, due to the limitation of computational power in our ultra-high-resolution simulation, we prescribed a two-step procedure for calculating collisionless dynamics, which is an $N$-body/hydrodynamic step and an $N$-body step.

The $N$-body/hydrodynamic step corresponds to the major star formation epoch, which is the first $2\times10^9$ years of our simulation run. In this step, since we mainly focus on the hydrodynamic response to the multiple SN explosions in an early epoch of galaxy formation, the gravitational potential is simply provided by the distribution of sub-galactic condensations having the Navarro-Frenk-White density profile. Therefore, the particle mass $m_j$ in the equation is replaced with $m_{\mathrm{NFW}}(|\mathbf{r}_i-\mathbf{r}_j|)$ that is given by the integration of equation (1) to a radius $|\mathbf{r}_i-\mathbf{r}_j|$. This procedure is 'Navarro-Frenk-White softening' in a manner similar to 'Plummer softening'.

As an $N$-body step, we simulate the long-term dynamical evolution of the model galaxy with an $N$-body scheme with one million particles to explore the quasi-equilibrium structure as a result of the pure stellar dynamical evolution. In this $N$-body step, we set up an $N$-body realization of each sub-galactic dark matter halo using $5\times10^4$ particles, and assign the initial positions and velocities to particles using the result of the $N$-body/hydrodynamic step. We treat totally $10^6$ dark matter particles and about $10^5$ *star particles*. The gravitational acceleration is simply calculated by equation (3)

### 3.2 Gas dynamics

The evolution of the gas is described by the three dimensional hydrodynamic equations for a perfect fluid in Cartesian geometry. The continuity equation, the momentum equation, and the energy equation can be written in a conservation form as

$$\frac{\partial \mathbf{U}}{\partial t} + \mathrm{div}\,\mathbf{F} = \mathbf{S}, \quad (4)$$

with

$$\mathbf{U} = \begin{pmatrix} \rho \\ \rho v_x \\ \rho v_y \\ \rho v_z \\ \rho e \end{pmatrix}, \mathbf{F}_x = \begin{pmatrix} \rho v_x \\ \rho v_x^2 + p \\ \rho v_x v_y \\ \rho v_x v_z \\ \rho v_x h_e \end{pmatrix}, \mathbf{F}_y = \begin{pmatrix} \rho v_y \\ \rho v_x v_y \\ \rho v_y^2 + p \\ \rho v_y v_z \\ \rho v_y h_e \end{pmatrix}, \mathbf{F}_z = \begin{pmatrix} \rho v_z \\ \rho v_x v_z \\ \rho v_y v_z \\ \rho v_z^2 + p \\ \rho v_z h_e \end{pmatrix}, \mathbf{S} = \begin{pmatrix} \dot{\rho}_s \\ \rho g_x \\ \rho g_y \\ \rho g_z \\ \rho \mathbf{v} \cdot \mathbf{g} - \Lambda + \Gamma_{\mathrm{SN}} \end{pmatrix}, \quad (5)$$

where $\mathbf{U}$ is a state vector of conserved quantities, $\mathbf{F}$ is the corresponding flux vectors, $\mathbf{S}$ is the source-term vector that includes sources and sinks of conserved quantities, such as heating and cooling terms and the gravitational acceleration. Here, $\rho$ is the gas density, $\mathbf{v} = (v_x, v_y, v_z)^T$ is the gas velocity vector,

$$e = \varepsilon + \frac{1}{2}|\mathbf{v}|^2 = h_e - \frac{p}{\rho}, \quad (6)$$

is the specific total energy, and $h_e$ is the enthalpy. The internal energy $\varepsilon$ is related to the gas pressure with the adiabatic index $\gamma (= 5/3)$ by

$$p = (\gamma - 1)\rho\varepsilon. \quad (7)$$

The mass transfer due to star-formation and SNe corresponds to $\dot{\rho}_s$, $\Lambda$ is the rate of radiative cooling, and $\Gamma_{\mathrm{SN}}$ is the rate of SNe heating, $\mathbf{g} = (g_x, g_y, g_z)^T$ is the gravitational acceleration of dark matter. As for baryonic components, the gravity is primarily exerted by dark matter in the early evolutionary stages and the hydrodynamic compression caused by SN explosions is dominant compared to the self-gravitational contraction in the later stages. Hence, we neglect the self-gravity of baryonic components in the *N*-body/hydrodynamic step. In the *N*-body step, dark matter and formed stars are treated as a totally self-gravitating system.

The set of hydrodynamic equations is solved by a finite volume scheme with Van-Leer-type flux-vector splitting scheme, which is based on the AUSM-DV described by Wada and Liou[61]. Liou and Steffen[62] developed a remarkably simple upwind flux vector splitting scheme called `advection upstream splitting method' (AUSM). It treats the convective and pressure terms of the flux function separately. The AUSM-DV has a blending form of AUSM and flux difference, and improves the robustness of AUSM in dealing with the collision of strong shocks. Furthermore, it has favourable properties of high-resolution for contact discontinuities and numerical efficiency. We extended it to second-order spatial accuracy using MUSCL[63] with a total variation diminishing limiter. Since this scheme has a great advantage due to the reduction of numerical viscosity, fluid interfaces are sharply preserved and small-scale features can be resolved as in the `piecewise parabolic method' (PPM) of Woodward and Colella[64]. The AUSM-DV scheme is, however, simpler and has a lower computational cost than PPM. The code is perfectly parallelized using the Message Passing Interface (MPI) and has passed successfully several tests and handles very well weak and high Mach number shocks.

### 3.3 Radiative cooling

The radiative cooling rates depend on the temperature and ionization states of the gas. Also, the chemical composition of the gas affects the cooling rate drastically. We calculated the emission properties of the gas component assuming an optically thin gas in collisional ionization equilibrium using MAPPINGIII code. MAPPINGS III is the successor of MAPPINGS II that is described by Sutherland and Dopita[65]. Using the hydrogen number density, the gas temperature, and the gas metallicity at each grid point, we calculate the spectral energy distribution (SED) of the gas for the wavelength range of 1–10,000,000 Å. In practice, to obtain the SED of the system we sum up the SED of each grid point for the gas components.

We should note that Mori, Umemura and Ferrara[5] demonstrated that Lyα extinction due to dust is negligible within the first $5\times10^7$ years because the metallicity of gas is very low. It is also the case for our simulation. But it is unclear for the later epoch. To know the escape fraction of Lyα photons, we need to calculate the three dimensional radiative transfer of Lyα emission with dust extinction. We realize that it is so challenging and important, but it is beyond the scope of this article. The full analysis including the contribution of dust extinction and emission will be given elsewhere.

### 3.4 Star formation

The star formation is allowed to proceed in the regions where the gas density exceeds the threshold density of 0.01 cm$^{-3}$ and the gas temperature is less than $10^4$ K. Once the gas is eligible to form stars, we assume that the local star formation rate is proportional to the local gas density and inversely proportional to the local free-fall time $t_{ff} = \sqrt{3\pi/(32G\rho)}$:

$$\frac{d\rho_*}{dt} = C_* \frac{\rho}{t_{ff}}, \tag{8}$$

where $C_*$ is a dimensionless star formation rate parameter. Using this expression for the star formation rate, the probability $p$ that a star is formed in a simulation time step $\Delta t$ is

$$p = 1 - \exp\left(-C_* \frac{\Delta t}{t_{\rm ff}}\right) \qquad (9)$$

A random number is then drawn to determine whether the star is formed during $\Delta t$. Once a region satisfies the star-formation criteria, we create new collisionless *star particles* there. It is noted that a *star particle* is not a single star but an assemblage of stars with an initial stellar mass function (IMF), which is described below. We prescribe that about one third of the gas mass in a cell forms a new collisionless *star particle*. The procedures are similar to those used by several authors[11, 16, 24, 27]. The subsequent motion of *star particles* is determined only by gravity. Here, we set the star formation rate parameter $C_*$=0.1 and note that the simulation is insensitive to the adopted value of this parameter[16, 27].

We compute the evolution of SED of *star particles* based on the method of stellar population synthesis, which utilizes the evolutionary tracks of stars with various masses and metallicities. Using the PÉGASE v2.0 code by Fioc and Rocca-Volmerange[66], we calculate the SED for 91–1,600,000 Å as a function of elapsed time from the formation of each *star particle*. The SED of a whole galaxy is then obtained by summing up SED's of all *star particles* ever formed at different times with different metallicities. The response functions for the *UBV* passbands are taken from Bessell[67] and those for the *K* passband from Bessell and Brett[68].

### 3.5 Supernova feedback

Massive stars are destined to explode as SNe and work as sources to transfer the energy, the synthesized heavy elements, and the materials (H and He) into interstellar medium. This feedback process is most critical in the simulation of galaxy formation. When a *star particle* (a stellar assemblage) is formed, stars more massive than $m_{\rm SN}$=8 $M_\odot$ are assumed to explode as Type II SNe (SNe II) with the explosion energy of $10^{51}$ ergs and eject synthesized heavy elements into interstellar medium leaving 1.4 $M_\odot$ remnants.

Here, as the initial stellar mass function (IMF), we assume Salpeter's IMF[69]. Then, given the slope index $x$ (=1.35) and lower and upper mass limits ($m_{\rm l}$, $m_{\rm u}$), the number of SNe II progenitors is calculated as

$$N_{SN} = \frac{x-1}{x} \frac{1-(m_{sn}/m_u)^{-x}}{1-(m_l/m_u)^{1-x}} \frac{m_s}{m_u}, \qquad (10)$$

where $m_s$ is the mass of a *star particle*, and $m_l$=0.1 $M_\odot$ and $m_u$=50 $M_\odot$ are assumed. These IMF parameters affect the heating rate of interstellar medium and the ejection rate of heavy elements from *star particles*. The number of SNe II is the most sensitive to the IMF slope. The total number of SNe II from a *star particle* can be related to the mass of ejected heavy elements from a *star particle* by making use of Table 2 of Tsujimoto *et al.*[70]. The mass of 2.4 $M_\odot$ of oxygen is ejected from a SNe II explosion. The released energy, heavy elements and materials are supplied to the eight cells surrounding the SN region. Then, the hydrodynamic evolution of heavy elements is followed by the same algorithm as the gas density.